\documentclass[sigconf]{acmart}
\AtBeginDocument{%
  }

\setcopyright{acmlicensed}
\copyrightyear{2018}
\acmYear{2018}
\acmDOI{XXXXXXX.XXXXXXX}
\acmConference[Conference acronym 'XX]{Make sure to enter the correct
  conference title from your rights confirmation email}{June 03--05,
  2018}{Woodstock, NY}

\usepackage{xspace} 
\usepackage{enumitem}
\usepackage{multirow}
\usepackage{graphicx} 
\usepackage{booktabs} 
\usepackage{changepage} 
 \newif\ifcomment\commenttrue

\ifcomment
\newcommand{\sumit}[1]{{\color{blue!70!black} [#1]$_\text{Sumit}$}}
\newcommand{\rini}[1]{{\color{magenta!70!black} [#1]$_\text{Rini}$}}
\newcommand{\sadra}[1]{{\color{red!70!black} [#1]$_\text{Sadra}$}}
\newcommand{\athena}[1]{{\color{lime!70!black} [#1]$_\text{Athena}$}}
\newcommand{\sujay}[1]{{\color{green!70!black} [#1]$_\text{Sujay}$}}
\newcommand{\esme}[1]{{\color{olive!70!black} [#1]$_\text{Esme}$}}
\newcommand{\run}[1]{{\color{purple!70!black} [#1]$_\text{Run}$}}

\else 
\newcommand\sumit[1]{}
\newcommand\rini[1]{}

\newcommand\sadra[1]{}
\newcommand\athena[1]{}
\newcommand\sujay[1]{}
\newcommand\esme[1]{}
\newcommand\run[1]{}
\fi

\newcommand{\Fone}{\textit{Decomposing Agent Execution}\xspace}
\newcommand{\Ftwo}{\textit{In-situ Explanations}\xspace}
\newcommand{\Fthree}{\textit{Surfacing Computation Logic}\xspace}
\newcommand{\Ffour}{\textit{Localized Editing}\xspace}
\newcommand{\Ffive}{\textit{Branching Exploration}\xspace}
\newcommand{\Fsix}{\textit{Scaffolded Task Formulation}\xspace}
\newcommand{\Fseven}{\textit{Scaffolding Question}\xspace}

\newcommand{\sheetcheck}{\textsc{Pista}\xspace}
\newcommand{\Sheetcheck}{\textsc{Pista}\xspace}

\usepackage{tikz}
\usepackage{xcolor}

\definecolor{myorange}{HTML}{FC9432}
\definecolor{mygreen}{HTML}{008A0E}
\definecolor{mypink}{HTML}{FF80DF}
\definecolor{mypurple}{HTML}{635DFF}
\definecolor{mybrown}{HTML}{82755B}

\DeclareRobustCommand{\cirfeat}[2]{%
  \smash{%
    \tikz[baseline=(char.base)]{
      \node[
        circle,
        inner sep=0.6pt,
        outer sep=0pt,
        fill=#1
      ] (char) {\color{white}{#2}};
    }%
  }%
}

\begin{document}


\title{Auditing and Controlling AI Agent Actions in Spreadsheets}
\settopmatter{authorsperrow=4}
\author{Sadra Sabouri}
\email{sabourih@usc.edu}
\affiliation{%
  \institution{University of Southern California}
  \city{Los Angeles}
  \country{USA}
}

\author{Zeinabsadat Saghi}
\email{saghi@usc.edu}
\affiliation{%
  \institution{University of Southern California}
  \city{Los Angeles}
  \country{USA}
}

\author{Run Huang}
\email{runhuang@usc.edu}
\affiliation{%
  \institution{University of Southern California}
  \city{Los Angeles}
  \country{USA}
}

\author{Sujay Maladi}
\email{maladi@usc.edu}
\affiliation{%
  \institution{University of Southern California}
  \city{Los Angeles}
  \country{USA}
}

\author{Esmeralda Eufracio}
\email{eeufracio@cpp.edu}
\affiliation{%
  \institution{California State Polytechnic University}
  \city{Pomona}
  \country{USA}
}

\author{Sumit Gulwani}
\email{sumitg@microsoft.com}
\affiliation{%
  \institution{Microsoft}
  \city{Redmond}
  \country{USA}
}

\author{Souti Chattopadhyay}
\email{schattop@usc.edu}
\affiliation{%
  \institution{University of Southern California}
  \city{Los Angeles}
  \country{USA}
}

\renewcommand{\shortauthors}{Sabouri et al.}

\begin{abstract}

Advances in AI agent capabilities have outpaced users' ability to meaningfully oversee their execution. AI agents can perform sophisticated, multi-step knowledge work autonomously from start to finish, yet this process remains effectively inaccessible during execution, often buried within large volumes of intermediate reasoning and outputs:
by the time users receive the output, all underlying decisions have already been made without their involvement. This lack of transparency leaves users unable to examine the agent's assumptions, identify errors before they propagate, or redirect execution when it deviates from their intent.
The stakes are particularly high in spreadsheet environments, where process and artifact are inseparable. Each decision the agent makes is recorded directly in cells that belong to and reflect on the user.
We introduce \sheetcheck, a spreadsheet AI agent that decomposes execution into auditable, controllable actions, providing users with visibility into the agent's decision-making process and the capacity to intervene at each step. A formative study (\(N = 8\)) and a within-subjects summative evaluation (\(N = 16\)) comparing \sheetcheck to a baseline agent demonstrated that active participation in execution influenced not only task outcomes but also users' comprehension of the task, their perception of the agent, and their sense of role within the workflow. Users identified their own intent reflected in the agent's actions, detected errors that post-hoc review would have failed to surface, and reported a sense of co-ownership over the resulting output. These findings indicate that meaningful human oversight of AI agents in knowledge work requires not improved post-hoc review mechanisms, but active participation in decisions as they are made.

\end{abstract}



\keywords{mixed-initiative systems, verification, agent, end-user, spreadsheet}

\begin{teaserfigure}
\centering
  \includegraphics[trim=0 6 0 6, clip, width=0.8\textwidth]{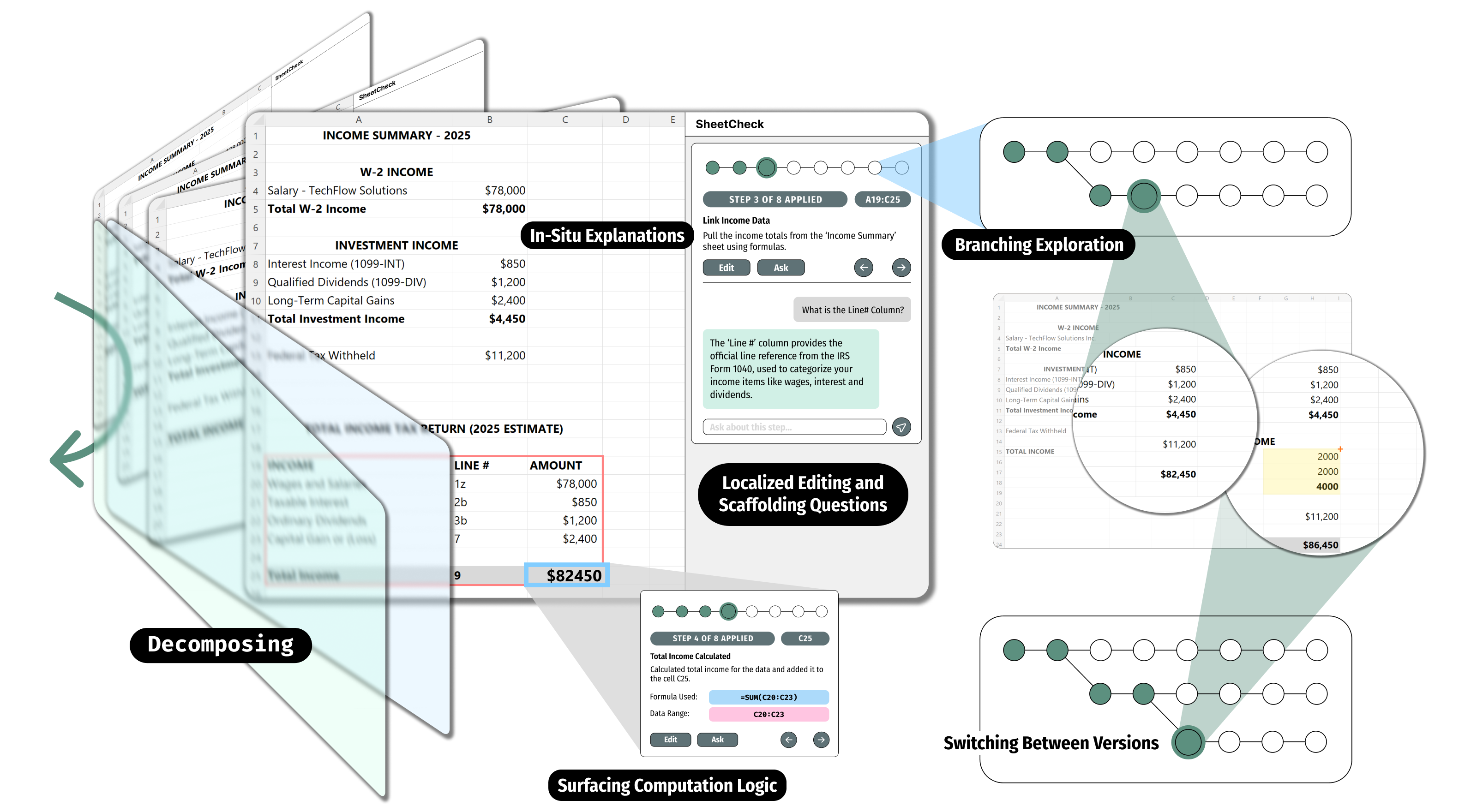}
  \caption{\sheetcheck is an AI agent for spreadsheets that \textit{decomposes} its execution into traceable, steerable steps, illustrated here on a tax return preparation task. At each step, \sheetcheck provides an \textit{In-Situ Explanations} of the action and surfaces the underlying formula and data range so users can verify correctness without inspecting individual cells. Users can probe the agent's reasoning through follow-up questions and issue targeted corrections scoped to the current step via \textit{Localized Editing}. When an edit alters computational logic that downstream steps depend on, \sheetcheck branches the execution path and regenerates subsequent steps from the edited state, preserving the original sequence so users can compare outcomes by switching between versions. This allows them to confidently explore alternative solutions.}
  \label{fig:teaser}
\end{teaserfigure}


\maketitle


\section{Introduction}

Artificial Intelligence (AI) agents are increasingly taking on complex, multi-step tasks across different knowledge work: analyzing datasets~\cite{hu2024infiagent}, writing code~\cite{eibl2025exploring}, managing documents~\cite{di2024streamlining}, and automating workflows that once required sustained human effort~\cite{anderson2023agency, pagliari2022human, Epperson2025Interactive, mozannar2025magentic}. As their capabilities expand, so does the scope of what they do on behalf of users, and the question of how humans and agents should share that work becomes more consequential. Yet in most current systems, agents work alone: they receive a goal, traverse a sequence of decisions, and return a result. The user is not present for any.

This absence may be workable when tasks are simple and errors are easy to spot. But as agents take on longer, more interdependent task sequences, writing formulas that drive downstream calculations, restructuring data in ways that break existing dependencies, and making assumptions that compound across steps, the gap between what the agent did and what the user can evaluate grows wider. The user receives a destination without having shared the journey. They must assess an artifact they did not witness being built, stand behind decisions they did not make, and detect errors in reasoning that were hidden from them.

This challenge is especially acute in structured knowledge work environments, such as spreadsheets. Spreadsheets are among the most widely used analytical tools across business, science, and government~\cite{Inala2024Data}, yet their computational logic is hidden behind a grid of values~\cite{rothermel1998you}: formulas encode assumptions about data sources, transformation rules, and dependencies that are invisible in the cells displaying their results. AI agents for spreadsheets, such as SheetCopilot~\cite{li2023sheetcopilot}, SheetMind~\cite{zhu2025sheetmind}, and TableTalk~\cite{liang2025tabletalk}, can automate complex multi-step tasks from natural language instructions, but currently achieve only 44--70\% success on real-world benchmarks~\cite{ma2024spreadsheetbench} and offer users little means to understand what the agent decided, why, or whether the result is correct. What makes spreadsheets a particularly sharp instantiation of this problem is that the artifact and the computation are the same object. In code agents, the running application is distinct from the code that produced it. In spreadsheets, there is no such separation: agent decisions are not stored elsewhere; they are the values and dependencies that constitute the artifact the user owns and is accountable for.

Prior work has addressed pieces of this problem. Gu et al.~\cite{Gu2023How} studied how analysts verify AI-generated data analyses, identifying procedural and data-oriented verification strategies and the cognitive burden verification places on users. Kazemitabaar et al.~\cite{Kazemitabaar2024Improving} showed that decomposing tasks into editable assumptions and steps improves steering and perceived control in data analysis notebooks. WaitGPT~\cite{xie2024waitgpt} visualized streaming code as an interactive diagram, helping users monitor and steer conversational data analysis agents in real time. Each of these contributions meaningfully reduces the gap between agent and user. Yet they share a common assumption: \emph{the user's role begins after the agent acts}. The human remains a reviewer of outputs, not a participant in their construction.

Reviewing a finished output, however, places the entire burden of oversight on a single moment of inspection~\cite{parasuraman2010, bowman2022, Gu2023How}. By that point, the agent's intermediate decisions have been compounded into a result whose surface gives little indication of the reasoning behind it. Errors legible at the step where they occurred become invisible in the aggregate. More fundamentally, the user's mental model of how the task should unfold no longer maps onto what the agent produces: rather than comparing their expectations against the agent's decisions as they develop, users must reconstruct the agent's logic from its outcomes. \textit{The question is not how to better present what an agent has already done, but how to involve the user in what the agent is doing, step by step, as it happens.}

To investigate this, we built \sheetcheck, from the Italian and Spanish for \textit{path} or \textit{trail}, and also \textit{clue}. The name captures the interaction model: rather than executing a task in a single pass and returning a finished result, \sheetcheck leaves a trail, decomposing agent execution into a sequence of traceable, steerable steps (Figure \ref{fig:teaser}). At each step, users see what the agent did and why, can probe its reasoning on demand, issue localized corrections, and branch into alternative execution paths, all before the next step proceeds. We instantiate this model as a Microsoft Excel add-in, a domain where the stakes of agent absence are especially high: every agent decision lands directly in the cells the user owns, is responsible for, and must stand behind. We first conducted a formative study ($N = 8$) to identify challenges with baseline spreadsheet agents, which surfaced five recurring challenges organized around the gulfs of evaluation, execution, and envisioning. Informed by these findings, we designed \sheetcheck around three design goals and seven features, then evaluated it in a within-subjects summative study ($N = 16$) comparing \sheetcheck to a baseline agent of equivalent capability.


Our findings show that being present during execution shapes not only what users can do but also how they understand the task, the agent, and their own role in the process. Users found the outcomes more comprehensible through interaction itself rather than in finished outputs, recognizing their own reasoning in the agent's steps. They caught errors they wouldn't have found in post-hoc review and felt co-authorship over the results they had helped construct. Taken together, these findings suggest that \sheetcheck is not just a transparent spreadsheet agent but an instance of a broader argument: that meaningful human oversight of AI agents in knowledge work requires participation in execution rather than taking on the role of a reviewer.



\section{Related Work}
AI agents for spreadsheets such as SheetCopilot~\cite{li2023sheetcopilot}, SheetMind~\cite{zhu2025sheetmind}, and TableTalk~\cite{liang2025tabletalk}
turn user instructions into sequences of spreadsheet actions, yet offer
little transparency into their decisions, leaving users with limited means
to verify outputs, understand agent reasoning, or steer toward desired
outcomes. These risks grow as automation increasingly displaces human
agency in knowledge work~\cite{fanni2023agency, pagliari2022human}. We
situate our work at the intersection of two lines of research: how
interfaces can support users to (1) interpret and verify AI agent outputs,
and (2) steer agent execution meaningfully.

\textbf{Tracing and Verifying AI Agent Outputs.}
Users of data analysis tools face high cognitive load when reviewing
AI-generated responses~\cite{khan2025cognitive, 10.1145/3654777.3676374}, often
feeling overwhelmed~\cite{Kazemitabaar2024Improving} and struggling to detect
misalignment between their intent and the agent's
output~\cite{Gu2023How, Kazemitabaar2024Improving}. These challenges
compound in multi-step pipelines where intermediate states are hidden:
users overlook critical errors even when the final outputs appear
stable~\cite{Inala2024Data}, and up to 27\% of agent failures involve
silent state integrity violations despite correct-looking
results~\cite{Rahman2025LLM-Based}. Recent tools address this by progressively
disclosing agent state~\cite{springer2019progressive}: WaitGPT transforms
streaming code into stepwise visual data operations with before/after table
states~\cite{10.1145/3654777.3676374}, ColDeco~\cite{ferdowsi2023coldeco}
decomposes LLM-generated formulas into intermediate helper columns that
surface edge cases, and BISCUIT~\cite{cheng2024biscuit} generates ephemeral
UI scaffolds in notebooks so users can explore alternatives before code is
finalized. These approaches share a common theme: making agent reasoning
legible enough for users to form accurate mental models of what changed
and why.

However, making reasoning visible does not automatically lead to
appropriate reliance. Bu{\c{c}}inca et al.~\cite{bucinca2021trust} showed that
cognitive forcing interventions reduce overreliance, but users preferred
the less demanding systems they performed worse with.
Vasconcelos et al.~\cite{vasconcelos2023explanations} formalized this
through a cost-benefit framework: overreliance drops when explanations make
verification cognitively tractable, but when the cost of engaging exceeds
the perceived benefit, users default to accepting AI output. These findings
suggest that post-hoc review of finished output is insufficient for
calibrated trust, and motivate an alternative: embedding verification
within the execution process itself. This is the approach Pista takes.

\textbf{Steering AI Agent Execution.}
The rapid expansion of AI capabilities has exacerbated concerns about
the loss of human agency~\cite{fanni2023agency} and diminishing user
engagement~\cite{10.1145/3654777.3676374} in automated workflows. A prerequisite
for effective steering is the ability to specify intent, yet this remains
a persistent bottleneck. Subramonyam et al.~\cite{subramonyam2024bridging}
coined the term ``gulf of envisioning" to describe the cognitive gap between
users' goals and their ability to formulate effective LLM prompts. 

Recent work has converged on design patterns that make
agent plans visible and editable. Cocoa exposes an interactive plan as a
list of steps users can reorder, rewrite, or delete~\cite{feng2024cocoa}.
Magentic-UI provides a natural-language plan that users can review and
modify before and during execution~\cite{mozannar2025magentic}. However, plan visibility alone is
not enough: He et al.~\cite{he2025planthenexecute} found that users can be
easily misled by plausible-sounding but flawed plans, highlighting the need
for verification affordances at each execution step.

In data analysis specifically, Kazemitabaar et
al.~\cite{Kazemitabaar2024Improving} showed that stepwise decomposition
interfaces significantly improve perceived control over global re-prompting.
SQLucid~\cite{tian2024sqlucid} demonstrated that letting users edit natural
language explanations of SQL queries improved task accuracy for non-experts.
Data Formulator~\cite{wang2025dataformulator} introduced branching history
for exploring alternative analytical directions without losing prior work.
Dango~\cite{chen2025dango} combined proactive clarification with editable
step-by-step explanations for data wrangling. Prior work on steerability
and verification has largely focused on code as the
medium~\cite{Kazemitabaar2024Improving, mcnutt2023notebooks}, exposing generated
code as an editable intermediate artifact through which users can inspect
and redirect agent behavior. We extend this to the spreadsheet setting, where
there is no such intermediate layer: agent decisions land directly in the
artifact the user owns, making the question of when and how to intervene
more imminent and consequential.

\section{Formative Study}

To better understand how users collaborate with spreadsheet agents and to motivate our design decisions, we conducted a formative study with eight participants ($N=8$).

\subsection{Technology Probe}

Traditional agents deliver complete answers in a single turn, hiding how they arrived at those answers. Similar agents in spreadsheets prevent users from seeing which columns the agent used, which rows it filtered, or why it made certain choices, leaving them unable to assess whether results match their intent, guide the agent when outputs diverge from their goals, or develop grounded expectations about what the agent will do next. To surface these challenges in a realistic context, we built a technology probe as a Microsoft Excel add-in with two core features.

\textbf{S1. Decomposing Agent Execution.} Rather than executing a task in a single pass, the probe decomposes the agent's plan into incremental steps. The agent first generates a complete plan of intended actions, then pauses at each step and advances only when the user clicks forward. Each step includes a plain-language description of what it does (e.g., ``Insert IFERROR month-over-month profit growth formula...'') and highlights all affected cells.

\textbf{S2. Localized Editing.} Users can intervene at any individual step without restarting the task. An Edit button opens a localized input field where users can provide natural-language corrections (e.g., ``use 3 decimal places for growth percentages''). The agent recalculates the affected step and propagates changes to subsequent steps, allowing incremental steering rather than full re-prompting.

\subsection{Setup}
\subsubsection{Protocol.}
Sessions were conducted remotely via Zoom. Participants watched a 1-minute overview video and installed the probe beforehand. At the start, we clarified that we were evaluating the tool, not participant performance. Each session included a 15-minute food warehouse restocking task, calculating order costs to sustain inventory given current stock and sales rates, with bonuses to engage early finishers. Task correctness and appropriateness were confirmed by a financial analyst. All participants completed the main task and at least one bonus. Sessions concluded with a 15-minute semi-structured interview, which was audio- and video-recorded with consent. The study was IRB-approved.

\subsubsection{Participants.}
We recruited participants via institutional mailing lists, university bulletin boards, and snowball sampling. Participants spanned multiple disciplines, including Computer Science ($n=4$), Electrical Engineering ($n=2$), Linguistics ($n=1$), and Political Science ($n=1$), reflecting varied technical backgrounds and spreadsheet use cases. The sample included five men and three women aged 26 to 30 ($M=27.5$, $SD=1.41$). Five participants had more than five years of experience working with spreadsheet environments, and six used spreadsheets daily, primarily for statistical analysis. All reported frequent use of large language models, indicating familiarity with AI-assisted tools.

\subsubsection{Analysis.}
We analyzed sessions using inductive coding. Two researchers independently annotated probe interactions and interview transcripts. After each coding session, they updated the codebook through negotiated agreement, reaching strong inter-rater reliability (Cohen’s $\kappa=0.91$).

\subsection{Findings}
Overall, participants responded positively to the core idea of decomposing agent execution into steps, praising it for making verification easier. At the same time, five challenges emerged that limited participants' ability to evaluate the agent's output, act on what they observed, and envision what alternatives were worth pursuing. To structure these challenges, we draw on three gulfs of interaction from the literature. The \textit{gulf of evaluation}~\cite{norman1988psychology} describes the gap in understanding what a system has done, the \textit{gulf of execution}~\cite{norman1988psychology}, captures the difficulty of translating intent into action, and \textit{gulf of envisioning}~\cite{10.1145/3613904.3642754} reflects the challenges in forming goals and imagining possibilities. We organize the challenges below by the gulf of interaction they primarily relate to.

\noindent \textbf{A. Gulf of Evaluation.}\\
\noindent\emph{[C1] Unable to probe and clarify the agent's decisions.} Participants appreciated the explanations alongside each step in execution, but found them often insufficient for understanding the agent's rationale. For example, FP7 noted that the agent did not explain \textit{``why it chose that [specific] option''} when \textit{``there were multiple options''}. Without such justification, participants struggled to assess whether the agent's choices were appropriate. Three participants desired a feature to \textit{``ask follow-up questions''} (FP5) to probe the agent's reasoning, noting that static explanations alone were not enough; they needed a way to actively engage with the agent's rationale to develop a deeper understanding of each step.

\noindent\emph{[C2] Lack of visibility into the underlying computation and data propagation.} The technology probe highlighted which cell values changed between steps, but participants found this insufficient for verification because it did not reveal the logic behind those changes. Four participants (FP1, FP3, FP6, FP8) noted that the computation logic was hidden behind computed values. While they could inspect formulas and cell references by clicking into cells, doing so across a large affected range was impractical. FP1 and FP7 suggested displaying formulas, references, and dependencies upfront so they could audit not just what changed, but how values were computed and how changes propagated.

\noindent\textbf{B. Gulf of Execution}~

\noindent\emph{[C3] Low confidence in editing and exploring alternatives.} The technology probe allowed participants to issue corrections scoped to a single step, which all participants found useful. However, five participants (FP1, FP2, FP6, FP7, FP8) remained hesitant to make edits because they could not predict how an edit would propagate and what \textit{``exactly is going to change in next steps''} (FP6), making even small changes feel risky. Additionally, participants complained that there was no way to compare alternatives or \textit{``revert a step''} (FP7), so they could not confidently explore other options without risking changes they could not reverse.

\noindent \textbf{C. Gulf of Envisioning}~

\noindent\emph{[C4] Difficulty Specifying Tasks and Constraints Upfront.} Three participants (FP4, FP5, FP6) struggled to translate their goals into concrete instructions. Particularly in the initial phase, they usually issued high-level prompts without specifying the assumptions, constraints, or standards that should govern how the task is carried out (FP4). Without support for articulating goals and constraints, their instructions remained underspecified, leaving key decisions to the agent that might not align with their intent.

\noindent\emph{[C5] Not knowing what to question and what else to try.} Participants were often unsure what aspects to scrutinize or how the result could be improved. All of them expressed a need for a \textit{``second pair of eyes''} that offers different perspectives from which the output could be verified, for instance, surfacing potential issues they had not considered. Several also wanted help surfacing \textit{``unknown unknowns''} (FP6) and exploring other plausible approaches, such as \textit{``browsing alternative options''} (FP4) or asking for follow-up suggestions (FP6) at each step.

\section{\sheetcheck}
\begin{figure*}[tbh]
    \centering
     \includegraphics[width=\linewidth]{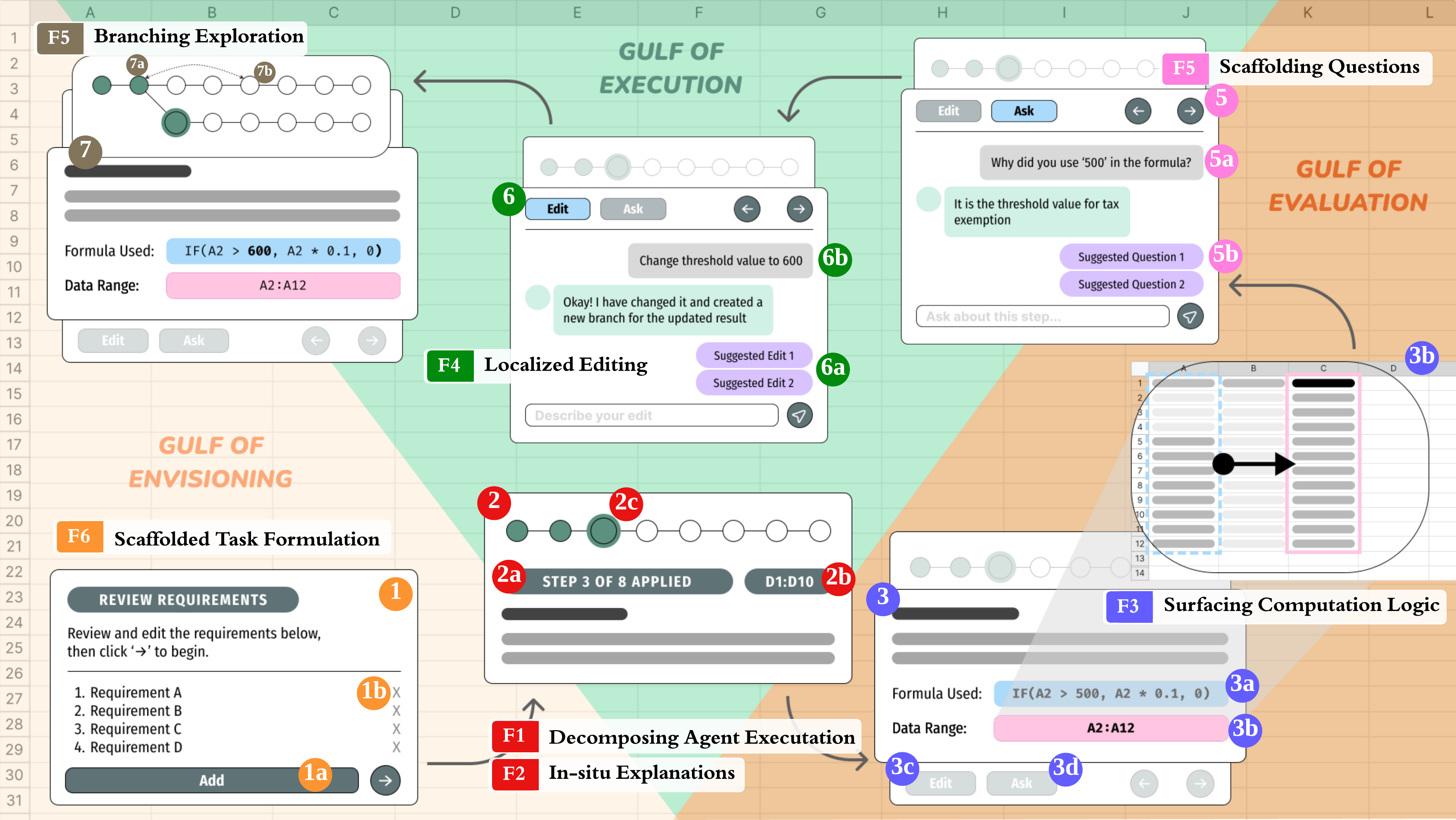}

    \caption{\textbf{\Sheetcheck decomposes AI agent execution into traceable, steerable steps.} It begins by generating modifiable requirements \cirfeat{orange}{1}, which users can refine by adding \cirfeat{orange}{1a} or removing \cirfeat{orange}{1b} items. \sheetcheck then situates each step \cirfeat{red}{2} within the plan \cirfeat{red}{2a}, highlights the affected spreadsheet range \cirfeat{red}{2b}, and shows progress via a node-based view \cirfeat{red}{2c} (current step highlighted). At each step, it displays the applied formula \cirfeat{mypurple}{3a}, highlights impacted data \cirfeat{mypurple}{3b}, and provides Ask \cirfeat{mypurple}{3d},\cirfeat{mypink}{5} and Edit \cirfeat{mypurple}{3c},\cirfeat{mygreen}{6} controls with scaffolded suggestions \cirfeat{mypink}{5a}, \cirfeat{mygreen}{6a}. Edits branch execution into a new plan \cirfeat{mybrown}{7}, and users can navigate between steps and branches \cirfeat{mybrown}{7a}, \cirfeat{mybrown}{7b}.}
    \label{fig:feature-figure}
\end{figure*}

We introduce \sheetcheck, a spreadsheet agent that executes in steps for users to review, steer, and explore alternatives along the way. \sheetcheck is implemented as an extension for Microsoft Excel that works on both the web and desktop versions. It appears as a side panel to the right of the normal spreadsheet workspace. Informed by the challenges observed in our formative study, we organize the design around three design goals (DG1-DG3). Figure~\ref{fig:feature-figure} outlines the workflow of using \sheetcheck and illustrates its core features.

\subsection{Making Agent Execution Traceable and Digestible \small{[DG1]}}
Participants in our formative study liked the idea of following the agent's execution step by step, but also raised a tension in how granular those steps should be. Decomposing the execution into too many fine-grained steps felt verbose and disrupted participants' workflow, while bundling too many operations into a single step made it difficult to interpret and verify. To navigate this tradeoff, \sheetcheck halts when the agent (a) modifies data or computational logic, e.g., writing formulas or adding derived columns, or (b) makes a decision that affects subsequent steps and outcomes, e.g., restructuring data into a pivot table versus aggregating with formulas. All low-level operations needed to carry out such a change or decision are bundled and presented as one step [F1].

A single step may propagate changes across many cells, and participants found it overwhelming when our technology probe highlighted every affected cell. Instead, they preferred to see how results were computed directly to understand the intent behind such changes [C2]. Accordingly, \sheetcheck pairs each step with an explanation of what the agent did and why [F2], and makes the underlying computation logic explicit by presenting formulas, dependencies, and affected data ranges right in the UI [F3]. For example, if the agent inserts a \texttt{SUMIF} formula, the panel shows the formula, the source range it reads from, and the output cells it writes to. This lets users quickly verify correctness by inspecting the computational logic rather than spot-checking hundreds of cells. Together, these designs aim to make each step digestible in scope [F1], grounded in a clear rationale [F2], and transparent for efficient verification [F3].

\subsection{Enabling Granular Intervention and Exploratory Refinement \small{[DG2]}}
Prior work has shown that users are more effective at refining AI-generated output when they can target specific segments rather than re-prompting the entire task, which risks overwriting parts that were already correct~\cite{Wang2023RePrompt:}. Following this principle, \sheetcheck allows users to issue follow-up instructions scoped to the current step, so the agent only revises changes introduced there while the rest of the spreadsheet remains intact [F4].

However, our formative study found that localized edits alone were not enough to make participants comfortable intervening, because they could not predict how a change would propagate to later steps [C3]. To address this, \sheetcheck supports branching [F5]. When an edit alters the computational logic or data that downstream steps depend on, \sheetcheck automatically creates a new branch in which all subsequent steps are regenerated from the edited state. Branches are displayed as a tree, and users can click any branch node to switch between execution sequences and compare how different edits lead to different outcomes. Because the original branch is always preserved, users can freely create further branches to try variations and only commit to a branch when they are satisfied.

Localized editing [F4] lets users make precise changes without affecting unrelated parts of the spreadsheet. Branching [F5] further supports users in actively exploring alternative solutions without worrying about losing prior work. By lowering the cognitive barrier to intervention, these features encourage users to iteratively refine the agent's output rather than passively accept it.

\subsection{Scaffolding Task Specification and Guided Exploration \small{[DG3]}}
Our formative study revealed two envisioning challenges: participants struggled to specify their preferences and requirements clearly, which left the agent to make implicit decisions that users only noticed and questioned after they were executed [C4], and when reviewing steps, they were often unsure what to scrutinize or what alternatives to consider [C5].

To help users communicate their intent before execution begins, \sheetcheck presents an overview of the actions it plans to take [F6], given the initial prompt and spreadsheet context. It also infers the users' intent and displays an editable list of decisions that the user may want to weigh in on, e.g., how to handle missing values. Users can reorder, add, or remove suggested specifications from this list and revise each in natural language. To scaffold users in identifying what to question and what alternatives are worth exploring during execution, \sheetcheck proactively suggests (a) follow-up questions beneath the explanation [F2] and (b) candidate edits within the editing input [F4], e.g., alternatives that the user might not have considered [F7]. Rather than requiring users to identify what to inspect or change, these suggestions make concrete opportunities for refinement visible at each step and guide users toward deeper engagement with the agent's output.

[F6] helps users externalize their intent more clearly before execution, while scaffolding suggestions [F7] within [F2] and [F4] help them identify opportunities for refinement as execution unfolds.


\subsection{Implementation Note}
\sheetcheck's front-end\footnote{\url{https://github.com/aceatusc/sheetcheck-ms-add-in}} is built in JavaScript and HTML. The back-end\footnote{\url{https://github.com/aceatusc/sheetcheck-backend}} is a Python Flask REST API that handles logging and routes the agentic calls to the Gemini LLM API, based on Microsoft Copilot for Excel agentic model, with output schemas defined as DSPy programs~\cite{khattab2023dspy} to ensure structured, consistent outputs.
Upon receiving a user request in a spreadsheet, the agent generates a sequence of Office.js code snippets. Snippets are transmitted to the front-end and executed via the Office.js API, with sheet state cached in the browser at each step to support \Ffive.

We used Gemini 3.0 Flash for all features requiring code generation and multi-part explanations (\Fone, \Ftwo, \Fthree, \Ffour, \Fsix), selected for its favorable balance of response latency and task accuracy. For \Fseven, where faster response time was the primary constraint and task complexity was lower, we used Gemini 3.1 Flash-Lite.

\section{Usability Study}
To evaluate the efficacy of \sheetcheck, we conducted a within-subject comparative usability study with $N=16$ participants. As a baseline, we stripped \sheetcheck of all features to simulate a typical spreadsheet AI agent of equivalent capability: one that applies changes directly to the spreadsheet and generates a summary of what was done. From this study, we answer the following research questions:

\begin{itemize}[leftmargin=*]
    \item \textbf{RQ1 (Traceability):} To what extent does \sheetcheck improve users' understanding of agent actions compared to a baseline agent?
    \item \textbf{RQ2 (Steerability):} How does \sheetcheck affect users' ability to detect, correct, and steer agent behavior toward their intended outcome?
\end{itemize}

\subsection{Participants}
We recruited 16 participants (ages 21–31, median 25.5) via mailing lists across university departments and regional professional networks in financial analytics, management, and computer science, supplemented by bulletin boards and snowball sampling. Inclusion required being over 18, English-speaking, and having prior Excel experience. Twelve were graduate students; the remainder were financial analytics professionals. Most had 1–5 years of spreadsheet experience; three had 10 or more. While 75\% used LLMs often or always, participants remained verification-oriented: 69\% manually double-check AI calculations and 75\% worry about missing errors they would have caught manually. When asked whether they would defer to an AI's statistical recommendation over their own analytical hunch, responses split evenly (8/16), reflecting diversity in how participants weigh AI authority against their own judgment.

\subsection{Protocol}

Each session began with a 10-minute onboarding task (planning a road trip from Los Angeles to New York) to familiarize participants with the tool's features. Participants then completed two counterbalanced tasks, one per condition. After each task, participants completed a post-task survey and interview, probing their understanding of agent actions, verification strategies, and steering experience. Treatment-specific questions probed feature usage. Sessions concluded with a comparative interview on verification effort, confidence, and ownership. Sessions were audio- and video-recorded with consent. Participants received a \$30 giftcard for the 1-hour session. The study was IRB-approved.

\subsection{Financial Analytic Tasks}
We selected two subtasks from SpreadsheetBench~\cite{ma2024spreadsheetbench}, a benchmark of real-world spreadsheet questions sourced from online Excel forums, and extended them to ensure participants would encounter both formula construction and data cleaning, the two most common daily challenges identified in the recruitment survey. Task A involved automating a final price-point column using a multi-product lookup formula, computing a cost distribution summary, and applying conditional color-coding. Task B required calculating weighted revenue via a cross-sheet probability lookup, aggregating results by priority, and handling case-insensitive deduplication. Each task was divided into five scored subtasks. A financial analytics expert confirmed that both tasks were unambiguous, equally difficult, self-contained, and could be completed within 15 minutes.

\subsection{Measure}
We measured participants' interaction with and understanding of the agent’s outputs through three proxies: (i) the depth of their verbal explanations, (ii) their self-reported experience via a post-study questionnaire, and (iii) their behavioral engagement captured through interaction logs.

\subsubsection{Depth of Understanding}
\label{sec:depth_understanding_measure}
We assessed participants' understanding of AI changes by analyzing verbal explanations collected after each task via a think-aloud prompt (``explain what the agent did and why'').
Following the verbal protocol analysis methods of \citet{ericsson2017protocol} and the self-explanation framework of \citet{chi1994eliciting}, we developed a domain-specific codebook to analyze spreadsheet-based explanations. We coded each segment along four dimensions adapted to the context of data analysis: 
mentioning a procedural step or outcome of the agent (PROC),
explaining mechanistically why a change happened (MECH),
mentioning a specific formula or logic construct (FORM),
expressing where the agent went wrong (ERR), and
specifying concrete column, cell, or value (SPEC).
Raters resolved disagreements through discussion; final scores represent the agreed-upon count.
The count of coded segments per explanation served as a proxy for the understanding depth score.

\subsubsection{Post-study Questionare.}
After each session, participants completed a post-study survey rating their experience with each system on a seven-point Likert scale. Questions covered overall helpfulness, sense of control, mental effort, ability to guide and understand the AI, error detection and correction, and the degree to which the AI's approach aligned with their own reasoning process.

\subsubsection{User Interaction Data.}
We collected participants’ activities to capture their interactions, including the number of issues detected by the user or the tool, the number of fixed issues, the number of prompting turns, and the frequency of feature use, such as clicking the Edit and Ask buttons. We also measured their success rate as a proportion of subtasks completed.

\subsection{Analysis}
\subsubsection{Quantitative}
We verified normality using the Shapiro-Wilk test before each statistical comparison and used nonparametric alternatives when normality was violated. For multiple comparisons on the same data, we applied the Bonferroni correction.

\subsubsection{Qualtiative}





We analyzed interview sessions and screen recordings to understand how participants worked with agents. Two researchers coded the anonymized transcripts, starting with open coding and then meeting to discuss disagreements. This process produced 42 distinct codes, which we organized through thematic analysis~\cite{clarke2017thematic} to address each research question.
To measure task performance, we tracked the number of requirements participants completed. Each task included five requirements. We calculated the success rate by dividing satisfied requirements by total requirements for each session.
We used screen recordings to assess whether \sheetcheck helped participants catch problems during collaboration. We counted the number of misalignments participants identified while working with the agent.
Interaction effort was measured via prompt count and prompt length (in words), two proxies for interaction intensity and context load based on prior work~\cite{chatterji2025people,petridis2024promptinfuser}.

\section{Findings}

\begin{figure*}[h]
    \centering
    \includegraphics[width=\linewidth]{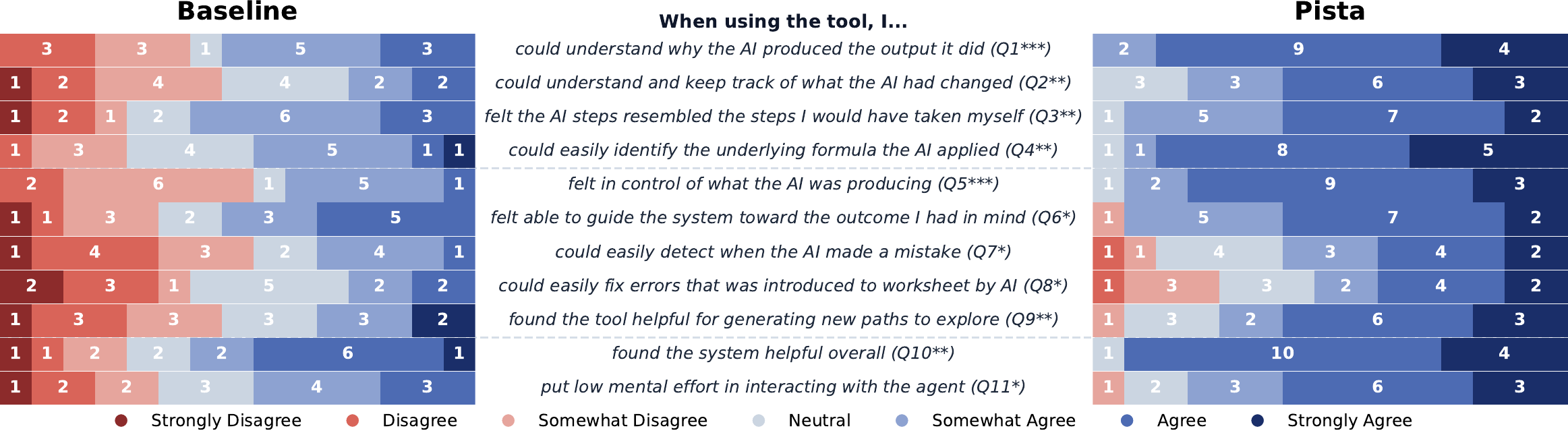}
    \caption{Likert ratings box plots for \sheetcheck vs. Baseline ($N = 15$). Significance via two-sided Wilcoxon signed-rank test: $^{*}p<.05$, $^{**}p<.01$, $^{***}p<.001$.}
    \label{fig:likert}
\end{figure*}

\begin{figure}
    \centering
    \includegraphics[width=0.7\linewidth]{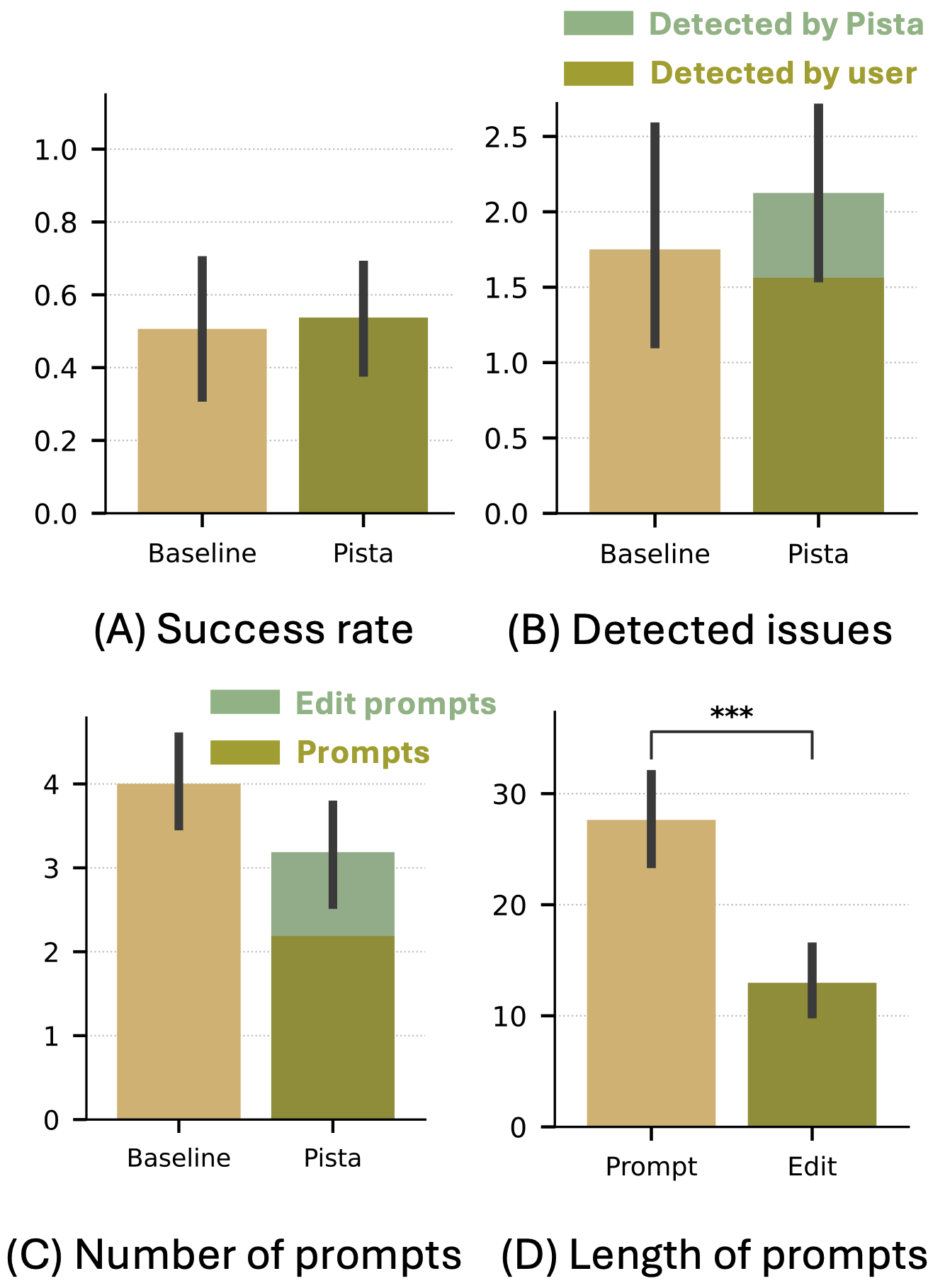}
    \caption{Per-session distributions across four measures: (A) task success rate, (B) issues detected (user- vs. tool-detected), (C) prompt count (standard vs. Edit), and (D) prompt length. Stars denote Mann-Whitney significance:  $^{***}p<.001$. Error bars show 95\% CIs.}
    \label{fig:performance}
\end{figure}

We begin by comparing how effectively participants understood and validated AI-generated spreadsheet solutions under \sheetcheck{} versus the Baseline. Both conditions achieved comparable success rates ($M_P = 0.53$, $SD_P = 0.31$ vs.\ $M_B = 0.50$, $SD_B = 0.40$; Figure~\ref{fig:performance}A), but \sheetcheck{} participants detected more issues per session ($M_P = 2.12$ vs.\ $M_B = 1.75$; Figure~\ref{fig:performance}B) while sending fewer prompts ($M_P = 3.18$ vs.\ $M_B = 4.00$; Figure~\ref{fig:performance}C) that were significantly shorter (Mann--Whitney $U = 3437.5$, $p < .001$; $M_P = 12.90$ vs.\ $M_B = 27.63$ words; Figure~\ref{fig:performance}D). Post-study ratings confirmed that participants found \sheetcheck{} significantly more helpful overall (Wilcoxon $W = 3.5$, $p < .01$; Figure~\ref{fig:likert}). Table~\ref{tab:feature_usage} summarizes feature usage. \Ffive was the most adopted feature (94\% of participants, median 3 uses), followed by \Ftwo (62\%). \Fsix, \Fseven, and \Fthree each reached roughly half or fewer participants. Qualitatively, participants described \sheetcheck{} as offering active presence and growing trust, while the Baseline was repeatedly characterized as a ``black box'' that delivered results without explanation.

\begin{table}[tbh]
\centering
\renewcommand{\arraystretch}{1}
\caption{\sheetcheck feature usage statistics: average (Avg), median (Med), maximum (Max) time used in a session, and number and percentage of participants used (n) across $N=16$.}
\label{tab:feature_usage}
\begin{tabular}{@{} l c c c r @{}}
\toprule
\textbf{Feature} & \textbf{Avg} & \textbf{Med} & \textbf{Max} & \textbf{n} (\%) \\
\midrule
\Ffive & 2.62 & 3   & 4 & 15 (94\%) \\
\Ftwo           & 1.23 & 1   & 3 & 10 (62\%) \\
\Fsix     & 0.60 & 1   & 1 & 9 (56\%)  \\
\Fthree            & 0.67 & 1 & 2 & 6 (38\%)  \\
\bottomrule
\end{tabular}
\end{table}

We now examine these patterns through each research question.

\subsection{RQ1: \sheetcheck{} Enhances User Comprehension of Agent Changes}

\sheetcheck's progressive pause-and-reveal shifted participants from reviewing completed spreadsheets post-hoc to tracing each formula as it was applied and reasoning about why it was applied. As shown in Figure~\ref{fig:likert}, participants using \sheetcheck{} reported they could better identify the agent's changes (Q1, $p < 0.001$), understand the rationale behind them, and track them more easily (Q2, $p < 0.01$).

\paragraph{Transparency.} Fourteen participants described \sheetcheck{} as making the agent's reasoning more visible than the Baseline. The stepwise execution (\Fone) let participants \textit{``exactly see what is happening''} (P11), which was especially valuable when the agent's logic involved unfamiliar concepts: P5 noted they could \textit{``check it in the middle''} even when they did not fully understand a step. Step-level explanations (\Ftwo) also helped participants catch issues they would have otherwise overlooked. P1, for instance, described how the walkthrough surfaced a conditional formatting decision they would not have thought to verify. Several participants (P2, P7, P8, P9, P11) reported that the \Fsix{} feature flagged discrepancies they would have missed entirely; as P2 noted, the correct output \textit{``would be zero if [they] didn't go back and see that.''} Beyond catching errors, participants found this transparency made the process educational: P2 described the tool as allowing them to \textit{``learn what [they]'re actually doing a bit more,''} in contrast to passively receiving a finished output. The Baseline, by comparison, offered no such foothold: P3 characterized it as \textit{``some black box model \ldots doing whatever it wants.''}

\begin{figure}
    \centering
    \includegraphics[width=\linewidth]{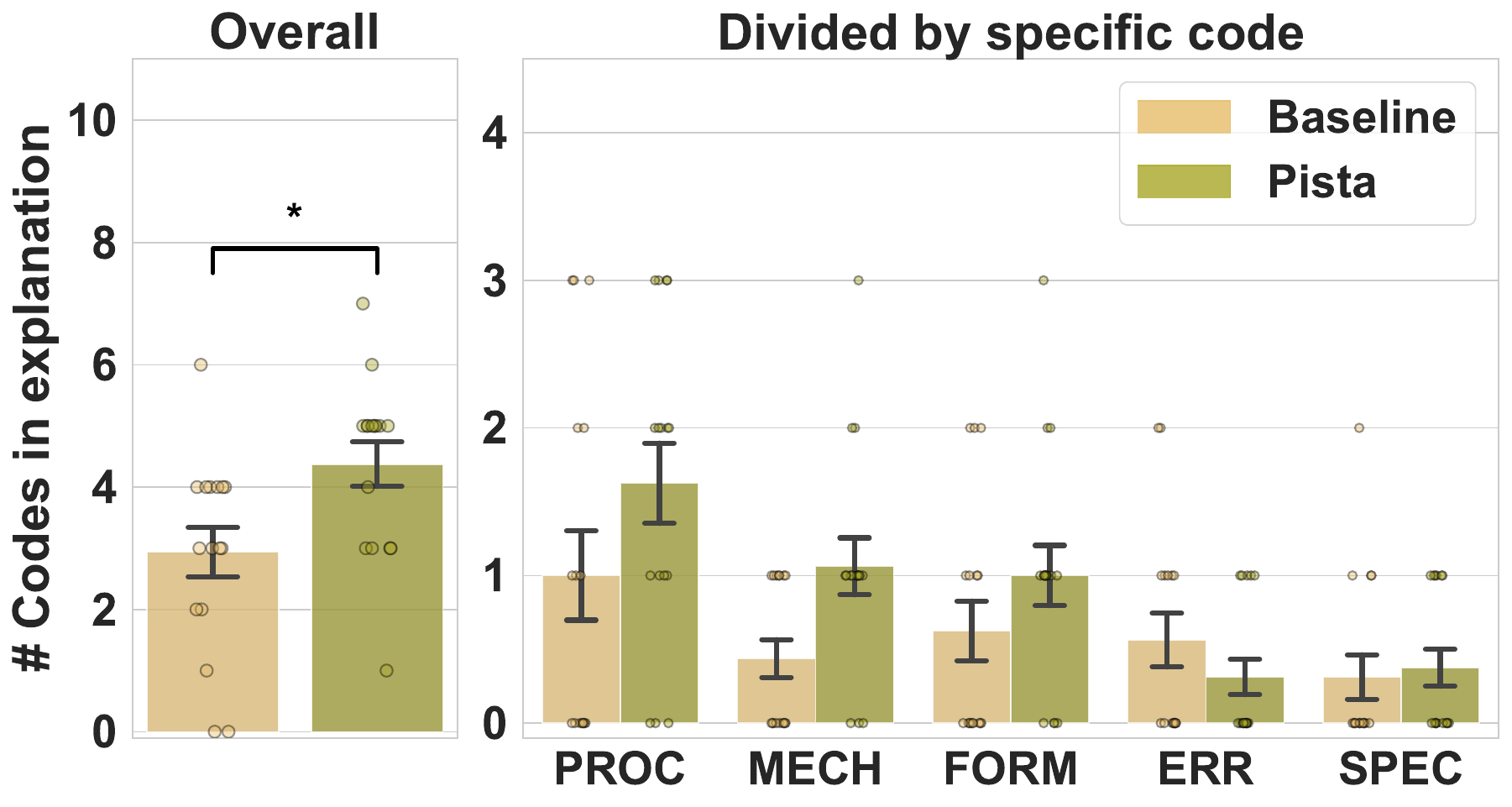}
    \caption{Number of qualitative codes in participants' explanations. \sheetcheck explanations include significantly higher codes than Baseline (left; $M=4.38, p=0.015$), a trend reflected across nearly all reasoning types (right). Error bars show the standard error of the mean.}
    \label{fig:understandint-code-count}
\end{figure}

\paragraph{Comprehension.} \sheetcheck's practice of naming and describing formulas at each step gave participants a semantic handle on changes: not just what cells were modified, but what the modification meant computationally. P6 articulated this: \textit{``you know exactly what the formulas are doing \ldots so it's just a matter of figuring out what the formatting issue is.''} Our post-study survey (Figure~\ref{fig:likert}) confirms that participants rated identifying formula changes (Q4) significantly easier with \sheetcheck{} than in the Baseline ($p < 0.01$). The \Fseven{} feature further supported comprehension by letting participants probe specific steps interactively, turning opaque outputs into ones they could interrogate on demand (P11).

\paragraph{Process Congruence.} Participants reported that \sheetcheck's step decomposition mirrored how they themselves would have approached the task (P4, P6, P8, P11). P6 noted: \textit{``\sheetcheck basically just gave me all the steps that I wanted to do before applying it \ldots I went ahead because it matched with what I wanted to do.''} P4 described a similar alignment at the level of language, observing that the agent \textit{``was just exactly translating my natural language prompt to Excel formulas.''} This congruence also aided verification: because participants held the same mental model as the agent's sequence, gaps between their expectations and the agent's actions were easier to spot mid-execution (P4, P6). Our post-study survey supports this: participants rated \sheetcheck's steps as significantly more resembling their own approach (Q3, $p < 0.01$; Figure~\ref{fig:likert}).

\paragraph{Confidence and Trust.} The transparency described above translated into greater trust. Participants consistently reported trusting \sheetcheck{} more because the reasoning behind each step was available for inspection (P1, P3, P13). P3 captured this directly: \textit{``I wasn't worried about the changes it's making, because it let me verify each step.''} This trust had behavioral consequences: P9 noted that trusting \sheetcheck{} more meant they \textit{``didn't feel the need to be as thorough''} in verification, a difference they called \textit{``a big deal in \ldots a rollout environment.''} Notably, trust was earned through the interaction itself rather than conferred on the output. As P7 remarked, \textit{``even though I was checking it manually, I still trusted it more,''} suggesting that the act of verification reinforced confidence rather than undermining it, consistent with~\citet{srinivasan2025adjust}.

These qualitative patterns are supported by our objective measure of understanding depth (see~\ref{sec:depth_understanding_measure}). Participants using \sheetcheck{} provided significantly richer explanations of the agent's process than those in the Baseline ($M_S = 4.38$ vs.\ $M_B = 2.94$; $p = 0.015$; Figure~\ref{fig:understandint-code-count}), covering more procedural steps, underlying mechanisms, and formula logic, while Baseline participants focused primarily on identifying errors.

\subsection{RQ2: \sheetcheck Supports Effective Steering and Error Correction}

Beyond the scaffolded pause-and-reveal, \sheetcheck made each step steerable, giving participants back agency and control. With the baseline, many described a reluctance to re-prompt, concerned that corrective instructions would cause the agent to redo everything and potentially break the spreadsheet. \sheetcheck's localized edit capability directly resolved this, allowing participants to explore making targeted edits at specific steps rather than re-running the full task, dramatically lowering the cost of intervention.
\paragraph{Ownership and Control.} As shown in Figure~\ref{fig:likert}, participants reported a greater sense of control over the AI in the \sheetcheck condition. This sense of control was closely tied to a feeling of co-authorship over the final output. Because participants reviewed and approved each step, the resulting spreadsheet felt like a collaborative product to them. P14 captured this well: ``\emph{I need to review every action and put my mental model in. I need to know what's right or wrong.}'' By contrast, the baseline left participants feeling like passive recipients: P3 described it as ``\emph{some black box model. I give it some context, and it's doing whatever it wants.}'' Twelve of 16 participants indicated that \sheetcheck gave them a stronger sense of ownership over the output.

\paragraph{Steerability.} Figure~\ref{fig:likert} shows that participants felt significantly more in control of guiding the agent toward their intended outcome with \sheetcheck. Participants noted that since each step was localized and inspectable, they could \textit{``pinpoint''} (P8) exactly where the agent's reasoning diverged from their intent and intervene to \textit{``correct [it]''} (P8) or \textit{``ask to do it in a different way.''} (P11). Conversely, with the Baseline agent, participants were forced to \textit{``re-write a full prompt to go through all the changes again''} (P3). \Sheetcheck influcend participatns efficiency as well. While performing at the same level, participants required significantly fewer prompts with \sheetcheck than baseline (($M_S=2.19$ vs. $M_B=4.00$), \(t(15)=4.53,p<0.001\)), substituting re-prompts with a single, targeted edit. Mann-Whitney test on the lengths of prompts (words) showed that \Ffour prompts were also significantly shorter than re-prompts ($M_S = 13$ vs. $M_B = 27.6$, $U=3437.5, p<0.001$), since with \Ffour feature, participants were able \textit{``to go back to a specific node and branch off from that''} (P9) as re-prompts required restating the full context. For example, during a session with the baseline agent, P7 prompted: \emph{Please change final price point (column R) with a formula to sum all of the item counts [sic] in a row, multiply by their prices (prices are in the Price sheet). Just in each row, consider items which are not blank!} However, when P7 used \Ffour, they simply targeted only the specific cells that needed to change and asked \sheetcheck to: \emph{apply the final price point formula for order 1 - 7}. Our findings suggest that direct, step-level manipulation enables more efficient steering without sacrificing output quality.

\paragraph{Structured exploration.} As shown in Figure~\ref{fig:likert}, participants also rated \sheetcheck as more supportive of exploring new approaches. The ability to pause at individual steps and redirect to a new branch (\Ffive) lowered the cost of experimentation. As P4 noted, the graph node navigation mechanism is valuable ``\emph{when you want to do things two different ways and still compare}'' the outcomes. This suggests that \sheetcheck's step-level control supports iterative refinement, enabling users to treat agent-assisted spreadsheet editing as an exploratory process rather than a one-shot transaction.

\section{Discussion}

Most AI assistance in productivity tools positions the agent as the executor and the user as a reviewer of finished outputs. Our findings \emph{challenge} this framing. When participants could engage with the agent's reasoning incrementally, inspecting, questioning, and redirecting at each step, they did not merely verify better: they thought differently about the task, about the agent, and about their own role in the process. \Sheetcheck{} is, in this sense, not just a more transparent agent but an argument for a different model of human-agent collaboration in knowledge work.

\subsection{Calibrated Trust Through Participation}

Conventionally, trust is a post-hoc judgment: the user receives an agent's output and decides whether to accept it. Our findings suggest that \sheetcheck{} disrupts this dynamic. Because participants reviewed and engaged with each step as it unfolded, trust was not a verdict rendered at the end, but a disposition that accumulated throughout the interaction. Participants reported trusting \sheetcheck{}'s output more, not because it was more accurate, but because they had been present, witnessing its construction.

This distinction matters for how we think about appropriate reliance on AI~\cite{lee2004trust}. The automation trust literature warns against both disuse, rejecting reliable AI output out of excessive skepticism, and misuse, accepting unreliable output without scrutiny~\cite{parasuraman1997}. \Sheetcheck{}'s interaction model creates a structural condition for calibrated trust by making the agent's process, not just its product, available for inspection. Participants caught errors through their own domain knowledge, and the \Fsix feature surfaced issues they would otherwise have missed~(P9). Critically, this scrutiny did not undermine trust: as our findings show that participants who checked more carefully trusted the output more, not less, consistent with recent work on verification-reinforced confidence~\cite{srinivasan2025adjust}.

This dynamic compounds with task complexity. Multiple participants noted that step-by-step control becomes more valuable as tasks grow more interdependent and error-prone, precisely the regime where automation bias~\cite{parasuraman2010} is hardest to resist. A system that exposes its process at high-stakes decision points gives users the evidence they need to trust selectively, directing their finite oversight budget~\cite{bowman2022} where it matters most.

\noindent\textbf{Design Implication.} Human-AI collaborative systems should aim to accrue trust through the interaction itself, not just through outcome quality. This means structuring agent execution where participants are present in the process, with sufficient visibility to form evidence-based judgments at each step. Attention-directing mechanisms that flag high-uncertainty or high-impact steps further support selective scrutiny without overwhelming the user.

\subsection{The Semantic Diff Primitive}

A recurring challenge in agentic systems is the mismatch between the
\textit{unit of change} and the \textit{unit of understanding}. Agents
typically surface what changed at the level of the artifact's surface:
highlighted cells, inserted lines of code, modified pixels. But users
reason about the operation that produced the change, the formula, the
transformation rule, the generative instruction, and whether that
operation matches their intent. When the surface unit and the semantic
unit diverge, verification becomes intractable: the user must mentally
reconstruct the agent's intent from its outputs, which is precisely the
cognitive burden that transparency is supposed to relieve~\cite{Kazemitabaar2024Improving, Gu2023How}.

We propose the \textit{semantic diff} as a design primitive that addresses
this mismatch. Rather than enumerating every affected element, a semantic
diff surfaces the computational unit driving the change, its logic,
its dependencies, and its scope, as the primary object of inspection.
In \sheetcheck{}, this means surfacing the formula applied to a column
rather than highlighting every cell it populates: if a VLOOKUP drives
500 changes, the user sees one formula and its explanation, not 500
highlights. Our summative study confirmed the value: participants
described formula-level visibility as what made verification tractable
rather than exhausting. The semantic diff generalizes across
artifact-producing agents: in code agents, the relevant unit is not the
line diff but the function or architectural decision being
applied~\cite{yan2025traces}; in document agents, it is the rhetorical
move, not the character-level edit. Building a semantic diff layer that
understands domain-specific artifact structure could serve as shared
infrastructure for inspectable agentic interfaces, making agent reasoning
legible not just to the immediate user but to anyone reviewing the
artifact afterward.

\noindent\textbf{Design Implication.} Agent interfaces operating on
structured artifacts should surface the \textit{semantic unit} of each
change, the formula, rule, or operation driving it, rather than
enumerating every affected element. This shifts verification from
auditing outputs to evaluating computational intent, and generalizes
across domains where the meaningful unit of change differs from the
visible unit.

\subsection{Scaffolding the Envisioning Gap}

\Sheetcheck{} directly addressed the gulf of
envisioning~\cite{subramonyam2024bridging}, the cognitive distance between
a user's goals and their ability to formulate a prompt that anticipates how
an LLM will interpret and execute it, through Scaffolded Task
Formulation~[F6] and Scaffolding Question~[F7]. Yet our findings reveal a
more fundamental tension: the envisioning gulf does not dissolve once a
task begins. It recurs at every step where the user must intervene.

This finding extends \citet{Kazemitabaar2024Improving}'s work in an important direction: 
They showed that interactive task decomposition reduces steering effort; our results suggest the residual burden falls on
\textit{formulating the correction}, not locating where to make it.
Morae~\cite{han2025morae} shows that proactive clarification at
high-uncertainty steps can surface this instruction gap before it blocks
the user. WaitGPT~\cite{10.1145/3654777.3676374} showed that visualizing
intermediate states helps users decide \textit{what} to change; our
findings suggest the next frontier is helping users decide \textit{how to
say} what they want to change. VRCopilot~\cite{vrcopilot2024} encountered
a related tension in immersive authoring, finding that scaffolded creation
modes helped users articulate the intent they could not express through
unconstrained natural language. While \sheetcheck{} addresses the
envisioning gap at task initiation~[F6] and execution~[F7], these
interventions assume users know their goals. When they do not, static
suggestions fall short, pointing toward responsive envisioning support that
adapts to uncertainty through structured examples~\cite{yan2025traces},
intent disambiguation~\cite{ambigchat2025}, and contrastive
explanation~\cite{wachter2017counterfactual}.

\noindent\textbf{Design Implication.} Agentic systems should treat
envisioning support not as a one-time affordance at task entry but as a
persistent layer throughout execution. At each step where a user might
intervene, the system should offer not just what the agent did and why,
but concrete, contextual starting points for expressing a correction,
such as exemplar edits, contrastive alternatives, or interactive
disambiguation, that lower the cost of formulating intent at the moment
it is needed most.

The interaction paradigm this work argues for, humans participating in
agent execution, rather than reviewing its outputs, is not specific to
spreadsheets. It applies wherever agents produce sequential, structured
artifacts and a human must remain meaningfully in the loop. The properties
that made this model effective here, trust accrued through participation,
semantic transparency at the level of computational intent, and persistent
envisioning support, generalize to any domain where the gap between what
an agent produces and what a user can evaluate is wide enough to matter.
In code agents, that gap is widening as natural language-driven generation
produces codebases users did not write~\cite{yan2025traces}. In research
synthesis, users discover what they need during review rather than
upfront~\cite{subramonyam2024bridging}. In document and design
agents~\cite{vrcopilot2024}, semantic diffs would surface rhetorical or
spatial operations rather than character-level edits. Across all these
domains, the central challenge is the one Pista addresses: making agent
reasoning legible enough for meaningful participation, without making
participation so costly that it defeats the purpose of having an agent
at all.

\section{Limitation}
Our participant pool skewed toward young, highly educated, and frequent LLM users; while we recruited from both professional and academic networks, this likely limits generalizability to less tech-savvy spreadsheet users.
Similarly, our tasks were scoped to financial analytics workflows, while the core interaction model is domain-agnostic; generalizability to other spreadsheet tasks and to longer, open-ended tasks remains an open question.
Our evaluation of steerability relies more on self-efficacy metrics than objective performance measures, in part because no established ground-truth metric exists for agent steerability from the human side. Prior efforts have focused mostly on the agent underlying capabilities to be steered, rather than interaction features for the user.

On the system side, step segmentation was developed around two principles: operations that modify data or computational logic, and decisions that affect subsequent steps and outcomes. Both studies confirmed the practical validity and perceived usefulness of this segmentation, though formal grounding in cognitive psychology theories of task decomposition remains an avenue for future work. Finally, alternative path suggestions and verification aspect scaffolding were constrained solely by system prompt instructions, a deliberate choice to explore the interaction space; future iterations should introduce more robust programmatic controls to better support trust and error recovery.

\section{Conclusion}

We presented \sheetcheck, a spreadsheet AI agent that decomposes execution into inspectable, steerable steps, giving users a path to follow through the agent's reasoning and a means to intervene at each step. Through a formative study ($N = 8$) and a within-subjects summative evaluation ($N = 16$), we found that participation in execution shaped not just
what users could do but how they understood the task, the agent, and their own role in the process: trust accumulated through the interaction rather than being conferred on outputs, users recognized their own reasoning in the agent's steps, and errors were caught that post-hoc review would have missed. These findings suggest that the central design challenge for agentic systems in knowledge work is not transparency alone, but presence: giving users the shared journey, not just the destination. 

\begin{acks}
We thank Prof. Jonathan May for helpful discussions during the early stages of this project, and the USC Information Sciences Institute (ISI) for providing the infrastructure that supported this research.
We are grateful to Prof. Jose Francisco Rubio, Amin Ahmadisharaf, and Maryam Ghareh Sheikhlou for their invaluable insights during the design and evaluation phases.
Finally, we extend our sincere appreciation to the students from the Departments of Financial Analytics, Management, and Computer Science at California State Polytechnic University, Pomona, and the University of Southern California, whose participation enabled the user studies.
\end{acks}

\bibliographystyle{ACM-Reference-Format}
\bibliography{main}

\appendix

\end{document}